\documentclass[5p,review,a4paper,onecolumn]{elsarticle}

\usepackage{lineno}
\usepackage{amsmath,graphicx}
\usepackage{amsfonts}
\usepackage{amssymb}
\usepackage{epstopdf}
\usepackage{xcolor}
\usepackage{array,booktabs}

\modulolinenumbers[5]

\journal{Signal Processing}

\usepackage{amsmath}
\usepackage{amsfonts}
\usepackage{amssymb}
\usepackage{graphicx}%
\usepackage{epstopdf}
\usepackage{ntheorem}
\usepackage{algorithm}
\usepackage{algorithmicx}
\usepackage{algpseudocode}
\usepackage{listings}
\usepackage{upquote}
\usepackage{xcolor}

\algnewcommand\algorithmicoutput{\textbf{Output:}} \algnewcommand\Output{\item[\algorithmicoutput]}
\algnewcommand\algorithmicinput{\textbf{Input:}} \algnewcommand\Input{\item[\algorithmicinput]}

\graphicspath{{./fig/}}

\begin{document}
	
	\begin{frontmatter}
		
		\title{On the Decomposition of Multivariate Nonstationary Multicomponent Signals}
		
		\author[mymainaddress]{Ljubi\v sa Stankovi\'{c}\corref{mycorrespondingauthor}}
		\ead{ljubisa@ac.me}
		
		\author[mymainaddress]{Milo\v s Brajovi\' c}
		\ead{milosb@ac.me}
		
		\author[mymainaddress]{Milo\v{s} Dakovi\'{c}}
		\ead{milos@ac.me}
		
			\author[mysecondaryaddress]{Danilo Mandic}
		\ead{d.mandic@imperial.ac.uk}
		
		\address[mymainaddress]{Faculty of Electrical Engineering, University of Montenegro, 81000 Podgorica, Montenegro}
		\address[mysecondaryaddress]{Imperial College London, London, United Kingdom}
		\cortext[mycorrespondingauthor]{Corresponding author}
		
		\begin{abstract}
With their ability to handle an increased amount of information, multivariate and multichannel signals can be used to solve problems normally not solvable with signals obtained from a single source. One such problem is the decomposition signals with several components whose domains of support significantly overlap in both the time and the frequency domain, including the joint time-frequency domain.  Initially, we proposed a solution to this problem based on the Wigner distribution of  multivariate signals, which requires the attenuation of the cross-terms.  In this paper, an advanced solution based on an eigenvalue analysis of the multivariate signal autocorrelation matrix, followed by their time-frequency concentration measure minimization, is presented. This analysis provides less restrictive conditions for the signal decomposition than in the case of Wigner distribution. The algorithm for the components separation is based on the concentration measures of the eigenvector  time-frequency representation, that are linear combinations of the overlapping signal components. With an increased number of sensors/channels a robustness of the decomposition process to additive noise is also achieved. The theory is supported by numerical examples. The required channel dissimilarity is statistically investigated  as well.    
  \end{abstract}


\end{frontmatter}

\section{Introduction}
	
It is well established that the use of the conventional Fourier analysis for the characterization and processing of signals with time-varying spectra is quite limited  \cite{dekompozicija0,dekompozicija_dsp,dekompozicija_meco,dekompozicija,dekompozicija2,dekompozicijaLFM,dekompozicija3,dekompozicija4,knjiga,VKLJS1,LJSMDTTB,mandic}. During the last few decades, these constraints have inspired the development of various powerful algorithms and approaches within the time-frequency signal analysis framework \cite{knjiga}.  

Traditional time-frequency analysis deals with univariate signals, frequently characterized through amplitude and frequency modulated oscillations \cite{knjiga}, \cite{mandic}. The short-time Fourier transform (STFT) and the Wigner distribution (WD) are commonly used time-frequency representations. In practice, signals are usually multicomponent, meaning that they can be represented as linear combinations of individual signals (components). Owning to its many desirable properties, Wigner distribution has been the basis of many instantaneous frequency (IF) estimators, developed to capture and describe frequency oscillations \cite{knjiga, VKLJS1,LJSMDTTB}. However, undesirable components, known as cross-terms, do appear in the Wigner distribution of multicomponent signals. With the intention to keep desirable properties of the STFT and high concentration of the WD, the S-method is developed as an alternative time-frequency representation, balancing between the previous two \cite{knjiga}.

For an independent characterization, each signal (component) in a multicomponent signal should be separated from others and individually analyzed \cite{dekompozicija,dekompozicija2,dekompozicijaLFM,dekompozicija3}. Such decomposition of multicomponent signals on individual components is possible for univariate signals by  means of the algorithm originally presented in \cite{dekompozicija}, which is based on the S-method. This type of decomposition is possible under the condition that time-frequency supports of individual components do not overlap. In the univariate case, in general, it is not possible to separate overlapped signal components, except for some very specific and a priori known/assumed signal forms, such as linear frequency modulated signals -- using chirplet transform \cite{dekompozicija_cirplet} or Radon transform \cite{dekompozicija_radon}, or sinusoidally modulated signals -- using inverse Radon transform \cite{iradon1}, \cite{iradon2}. 

Recently, new perspectives for the multicomponent signal decomposition have appeared, in light of the multivariate signal paradigm \cite{dekompozicija0}. Multivariate (multichannel) data have been largely available lately, as a result of new developments in the sensor technology. With the aim to exploit multichannel signal interdependence through a joint time-frequency analysis, concepts of modulated bivariate and trivariate data oscillations appeared  first, followed by the generalization of the concept to an arbitrary number of channels \cite{mandic,TSPMV,boashmv,bivariate}. The joint IF concept has been proposed in \cite{TSPMV}, as a characterization of multichannel data obtained by capturing combined frequency in all individual channels. The IF of a multivariate signal is defined as a weighted average of the IFs in all individual channels. Following the foundations of these basic time-frequency concepts for the multichannel data, synchrosqueezed transform has been reintroduced within the multivariate framework  \cite{mandic}. Furthermore, the wavelet ridge algorithm, as a tool for the extraction of local oscillatory dynamics of multivariate signal, is also defined for multivariate signals \cite{TSPMV}. Within the multivariate framework, significant research has also been conducted with the aim to place the empirical mode decomposition within the multivariate context \cite{emd1}-\cite{emd5}. Interestingly enough, this type of decomposition is possible only in the case of components which do not overlap in the time-frequency plane, even in the multivariate case.

Multivariate Wigner distribution has been the basis of the recently proposed approach for the decomposition of multivariate multicomponent signals \cite{dekompozicija0}. Exploiting the  significant reduction of undesirable cross-terms due to the multichannel signal nature, this method provides the possibility to efficiently extract the components with  overlapped  supports in the time-frequency domain, something that was not, in general, possible for univariate signals, using any known decomposition procedure. In particular, the autocorrelation matrix of Wigner distribution  is decomposed into eigenvectors. Using a steepest descent approach \cite{dekompozicija0}, they are linearly combined to form the extracted components. Besides the possibility to separate overlapped components, it has been even possible to apply the decomposition procedure to extract the IF of real-valued multichannel signals with amplitude variations proportional to phase variations \cite{dekompozicija_dsp}. The influence of channel phase differences (in the bivariate case) is analyzed in \cite{dekompozicija_meco}.

In this paper, the decomposition procedure is performed starting directly from a realization of signal autocorrelation matrix. This leads to less restrictive signal decomposition conditions, compared to the case of the decomposition based on multivariate Wigner distribution. It is shown that the eigenvectors of the analyzed matrix contains linear combinations of components overlapped in the time-frequency plane. These components are then extracted by minimizing the concentration of the linear combinations of eigenvectors. Numerical results verify the presented theory, with a special emphasis on robustness in noisy conditions and its relation to the number of channels. Overlapped components appear in various signal processing applications, such as in radar signal processing \cite{dekompozicija0}, multiple antenna systems \cite{R1}, some biomedical signals etc, to mention but a few.

{\color{red}
}

The paper organization is as follows. After a short overview of the background theory and basic definitions, Section \ref{sectionMMS} continues with the detailed analysis of multivariate multicomponent signals. In this section, the attention is devoted to the eigenvectors of signal autocorrelation matrix and their relations with signal components. Section \ref{decomposition_principle} presents the multivariate multicomponent signal decomposition approach, founded on the minimization of the concentration measure. Numerical results are given in Section \ref{examples}, while the paper ends with concluding remarks.


\section{Multivariate Multicomponent Signals}
\label{sectionMMS}
Discrete-time signals of the form%
\begin{equation}
\mathbf{x}(n)=
\begin{bmatrix}
a_1(n)e^{j\phi_1(n)}\\
a_2(n)e^{j\phi_2(n)}\\
\vdots \\
a_S(n)e^{j\phi_S(n)}
\end{bmatrix}, \,\,n=1,2,\dots,N,
\label{MVSIG}
\end{equation}
obtained by measuring a complex-valued signal $x(n)$ with $S$ sensors, are known as complex multivariate signals. The amplitude and phase of the original signal are modified by each sensor, to give $a_i(n)\exp(j\phi_i(n))=\alpha_ix(n)\exp(j\varphi_i)$.
In the case of real-valued measured signal, its analytic extension 
$$x(n)=x_R(n)+j\mathcal{H}\{x_R(n)\}$$ is commonly used, with $x_R(n)$ being the real-valued measured signal and $\mathcal{H}\{x_R(n)\}$ its Hilbert transform.
The analytic signal contains only nonnegative frequencies and the real-valued counterpart can be reconstructed. This form of signal is especially important in the instantaneous frequency interpretation within the time-frequency moments framework. 

Consider a multivariate signal obtained by sensing a monocomponent signal of the form $x(n)=A(n)\exp(j\psi(n))$. The value of this signal measured at a sensor $i$ can be written as $$a_i(n)\exp(j\phi_i(n))=\alpha_i\exp(j\varphi_i)x(n).$$ 
A real-valued form of this multivariate signal takes the form $a_i(n)\cos(\phi_i(n))$. According to  Bedrosian's
product theorem \cite{BB}, the complex analytic signal  $a_i(n)\exp(j\phi_i(n))=a_i(n)\cos(\phi_i(n))+j\mathcal{H}\{a_i(n)\cos(\phi_i(n))\}$
is a valid representation of the real amplitude-phase signal $a_i(n)\cos(\phi_i(n))$ if the spectrum of $a_i(n)$ is nonzero only within the frequency range $|\omega|<B
$ and the spectrum of $\cos(\phi_i(n)) $ occupies an nonoverlapping (much) higher frequency range. A signal is monocomponent if $a_i(n)$
is  slow-varying as compared to $\phi_i(n)$ variations. The signal model with slow amplitude variations, as compared to the phase variations, has been often considered in literature \cite{gabor,man2,man3,man4,man5,man6,man7}.

However, in general, for the case of multicomponent signals, the components are localized along more than one instantaneous frequency.

\subsection{Multivariate and Multicomponent Signals}
Consider a multicomponent discrete-time signal
\begin{equation}
x(n)= \sum_{p=1}^P x_p(n),
\end{equation}
with $P$ components of the form
\begin{equation}
x_p(n)=A_p(n) e^{j\psi_p(n)},
\end{equation}
where the component amplitudes $A_p(n)$ have a slow-varying dynamics as compared to the variations of the phases $\psi_p(n)$. Assume that components are independent signals, i.e., that no component can be written as a linear combination of other components (for all considered time instants $n$).
The corresponding multivariate signal is
then given by
\begin{equation}
\mathbf{x}(n)=
\begin{bmatrix}
\sum_{p=1}^P\alpha_{1p}x_p(n)e^{j\varphi_{1p}}\\
\sum_{p=1}^P\alpha_{2p}x_p(n)e^{j\varphi_{2p}}\\ 
\vdots \\
\sum_{p=1}^P\alpha_{Sp}x_p(n)e^{j\varphi_{Sp}}
\end{bmatrix}=\begin{bmatrix}
x^{(1)}(n)\\
x^{(2)}(n)\\
\vdots \\
x^{(S)}(n)
\end{bmatrix}.
\label{MVmcSIG}
\end{equation}
Signal in the $m$-th channel, denoted by $x^{(m)}(n)$, is obtained as a linear combination of the signal components $x_{p}(n)$ multiplied with complex constants $a_{mp}=\alpha_{mp} e^{j\varphi_{mp}}$, $m=1,2,\dots,S$, $p=1,2,\dots,P$, to give
\begin{equation}
\begin{bmatrix}
x^{(1)}(n)\\
x^{(2)}(n)\\
\vdots \\
x^{(S)}(n)
\end{bmatrix}=
\begin{bmatrix}
a_{11} & a_{12} & \dots &  a_{1P} \\
a_{21} & a_{22} & \dots &  a_{2P} \\
\vdots & \vdots & \ddots & \vdots \\
a_{S1} & a_{S2} & \dots &  a_{SP}
\end{bmatrix}
\begin{bmatrix}
x_1(n)\\
x_2(n)\\
\vdots \\
x_P(n)
\end{bmatrix}.  
\label{MVmcSIG2}
\end{equation}
We will introduce the notation
$$\mathbf{A}=\begin{bmatrix}
a_{11} & a_{12} & \dots &  a_{1P} \\
a_{21} & a_{22} & \dots &  a_{2P} \\
\vdots & \vdots & \ddots & \vdots \\
a_{S1} & a_{S2} & \dots &  a_{SP}
\end{bmatrix}
$$
for the $S\times P$ matrix that transforms the signal components to the measured signal.
 
\textit{Observation:} The maximum number $M$ of independent channels $x^{(1)}(n)$, $x^{(2)}(n), \dots$, $x^{(S)}(n)$ in $\mathbf{x}(n)$ is  
\begin{equation}
M=\min\{S,P\}.
\end{equation}  

The proof is evident since the transformation matrix in (\ref{MVmcSIG}) is an $S\times P$ matrix with $\mathrm{rank}\{\mathbf{A}\} \le \min\{S,P\}$.  

Note that if $S<P$ the maximum number of independent channels $x^{(1)}(n)$, $x^{(2)}(n), \dots$, $x^{(S)}(n)$ is equal to the number of sensors $S$, while if $S \ge P$ the maximum number of independent channels is equal to the number of components $P$.  

A matrix form of the previous relation between signals measured on $S$ sensors and $P$ signal components is
\begin{equation}
\begin{bmatrix}
x^{(1)}(1) & \dots & x^{(1)}(N) \\
x^{(2)}(1)& \dots & x^{(2)}(N) \\
\vdots & \ddots& \vdots \\
x^{(S)}(1) & \dots & x^{(S)}(N)
\end{bmatrix}=
\mathbf{A}\begin{bmatrix}
x_1(1) & \dots & x_1(N) \\
x_2(1) & \dots & x_1(N) \\
\vdots & \ddots& \vdots \\
x_P(1)  & \dots & x_1(N)
\end{bmatrix}.  
\label{MVmcSIG2M}
\end{equation}
or 
$$\mathbf{X}_{sen}=\mathbf{A}\mathbf{X}_{com}$$
where $\mathbf{X}_{sen}$ is an $S\times N$  matrix of sensed signal values with elements $x^{(s)}(n)$ and $\mathbf{X}_{com}$ is a $P\times N$  matrix of signal component samples with elements $x_p(n)$.
  
The autocorrelation matrix $\mathbf{R}$ of the sensed signal is defined by 
\begin{align}
\mathbf{R}= \mathbf{X}_{sen}^H\mathbf{X}_{sen},\label{rdef}
\end{align}
where $(\cdot)^{H}$ denotes the Hermitian transpose.
The elements of this matrix are
\begin{align} \label{relem}
R(n_{1},n_{2}) =\mathbf{x}^H(n_{2})\mathbf{x}(n_1)  =\sum_{i=1}^Sx^{(i)*}(n_2)x^{(i)}(n_1),
\end{align}
where $\mathbf{x}(n_{1})=[x^{(1)}(n_1) \,\, x^{(2)}(n_1)\,\, \dots \,\, x^{(S)}(n_1)]^T$ is the column vector of sensed values at a given instant $n_1$. 

The matrix $\mathbf{R}$  can be used for the analysis and characterization of multicomponent multivariate signals. It is also the starting point of the decomposition algorithm for multicomponent signals presented in this paper. 

Note that the sensed values $\mathbf{x}(n_{1})$ are the linear combinations of the signal components. Although the decomposition could be performed directly, based on the sensed signals, it would not be  computationally efficient for $S>P$ that is case common in the analysis. The efficiency is improved using the matrix eigen-decomposition of the autocorrelation matrix  $\mathbf{R}$. Some properties of this decomposition, needed for the analysis of multicomponent signals, will be reviewed next.

\subsection{Eigenvectors and Linear Combination of Vectors}
For any square matrix, the eigenvalue decomposition of a $K\times K$ dimensional matrix $\mathbf{R}$ gives 
\begin{equation}
\mathbf{R=Q}\mathbf{\Lambda}\mathbf{Q}^{H}=\sum_{p=1}^{K}\lambda_{p}\mathbf{q}%
_{p}\mathbf{q}_{p}^{H}, \label{eig}%
\end{equation}
where $\lambda_{p}$ are the eigenvalues and $\mathbf{q}_{p}$ are the corresponding eigenvectors
of $\mathbf{R}$.  Matrix $\mathbf{\Lambda}$ is a diagonal matrix with eigenvalues $\lambda_{p},~p=1,\dots,K$ on the main diagonal whereas the matrix $\mathbf{Q}$ is formed from eigenvectors as $\mathbf{Q}=\left[\mathbf{q}_1,\dots,\mathbf{q}_K\right]$. Note that the eigenvectors $\mathbf{q}_{p}$ are orthonormal.  

\textit{Remark 1:} Consider a set of nonorthogonal vectors $\mathbf{v}_m$, $m=1,2,\dots,M$. If a matrix $\mathbf{R}$ is defined by  
\begin{equation}
\mathbf{R}=\sum_{m=1}^M \mathbf{v}_m\mathbf{v}_m^H, \label{defRwv}
\end{equation}
then finding the eigenvectors of this matrix 
can be considered as the process of the orthogonalization of the space defined by vectors $\mathbf{v}_m$ whose energies are $\left\Vert \mathbf{v}_m \right\Vert_2^2=e_m$. 

Note this particular form of matrix is obtained for the multicomponent multivariate overlapping signals, since the elements of matrix $\mathbf{R}$ in (\ref{rdef}) are calculated as $R(n_{1},n_{2})  =\mathbf{x}^H(n_{2})\mathbf{x}(n_1)$.

The previous remark will be illustrated considering the cases with $M=1$, $M=2$, and an arbitrary $M$.
\begin{itemize}
\item 
If $M=1$ then the orthogonalization over $\mathbf{v}_1$ is not needed. In this case, the eigenvector of matrix $\mathbf{R}$ $\mathbf{q}_1=\mathbf{v}_1/\sqrt{e_1}$. This case appears exactly if the Wigner distribution is used in univariate signals. This property is used in the synthesis of signals with a given Wigner distribution.  

\item 
For $M=2$, the orthogonalization of the space defined by $\mathbf{v}_1$ and $\mathbf{v}_2$ is performed. In this case, the eigenvectors, $\mathbf{q}_{1}$, $\mathbf{q}_{2}$, as the orthogonal vectors over this space, can be written as two linear combinations of $\mathbf{v}_1$ and $\mathbf{v}_2$, that define matrix $\mathbf{R}$ in (\ref{defRwv}), that is
\begin{align*}
\mathbf{q}_{1}&=\gamma_{11}\mathbf{v}_1+\gamma_{21}\mathbf{v}_2\\
\mathbf{q}_{2}&=\gamma_{12}\mathbf{v}_1+\gamma_{22}\mathbf{v}_2.
\end{align*}

In order two prove this property we will start from definition (\ref{defRwv})
$$\mathbf{R}=\mathbf{v}_1\mathbf{v}_1^H+\mathbf{v}_2\mathbf{v}_2^H.$$
We assumed that the eigenvector $\mathbf{q}_{1}$ is of the form $\mathbf{q}_{1}=\gamma_{11}\mathbf{v}_1+\gamma_{21}\mathbf{v}_2$. The eigenvector of matrix $\mathbf{R}$ satisfies the relation $\mathbf{R}\mathbf{q}_{1}=\lambda_1\mathbf{q}_{1}$. Since
\begin{align*}
\mathbf{R}\mathbf{q}_{1}&=(\mathbf{v}_1\mathbf{v}_1^H+\mathbf{v}_2\mathbf{v}_2^H)(\gamma_{11}\mathbf{v}_1+\gamma_{21}\mathbf{v}_2)\\
&=\mathbf{v}_1\mathbf{v}_1^H(\gamma_{11}\mathbf{v}_1+\gamma_{21}\mathbf{v}_2)+\mathbf{v}_2\mathbf{v}_2^H(\gamma_{11}\mathbf{v}_1+\gamma_{21}\mathbf{v}_2)\\
&=\mathbf{v}_1(\gamma_{11}e_1+\gamma_{21}b_{12})+\mathbf{v}_2(\gamma_{11}b_{12}^*+\gamma_{21}e_2)
\end{align*}
where $b_{12}=\mathbf{v}_1^H\mathbf{v}_2$, we can obtain a system 
\begin{align*}\lambda_1 \mathbf{q}_1&=\lambda_1(\gamma_{11}\mathbf{v}_1+\gamma_{21}\mathbf{v}_2)\\
&=\mathbf{v}_1(\gamma_{11}e_1+\gamma_{21}b_{12})+\mathbf{v}_2(\gamma_{11}b_{12}^*+\gamma_{21}e_2).\end{align*}
 From this system of equations we can find $\gamma_{11}$, $\gamma_{21}$, and $\lambda_1$, based on $e_1$, $e_2$, and $b_{12}$ with  additional condition that $\left\Vert \mathbf{q}_1 \right\Vert_2^2=1$. The same holds for $\mathbf{q}_2$.

\item
This proof can be generalized for any $M$.
\begin{align*}
\mathbf{R}\mathbf{q}_i&=\sum_{m=1}^M \mathbf{v}_m\mathbf{v}_m^H\sum_{l=1}^M\gamma_{li}\mathbf{v}_l=\sum_{m=1}^M \mathbf{v}_m\sum_{l=1}^M\gamma_{li}\mathbf{v}_m^H\mathbf{v}_l\\&=\sum_{m=1}^M \mathbf{v}_m\sum_{l=1}^M\gamma_{li}b_{ml}=\sum_{m=1}^M \mathbf{v}_m B_{mi}
\end{align*}

From this relation and 
$$\lambda_i\mathbf{q}_i=\sum_{m=1}^M \mathbf{v}_m \lambda_i \gamma_{mi}$$
with $\mathbf{R}\mathbf{q}_{i}=\lambda_1\mathbf{q}_{i}$ follows the system
$$\sum_{m=1}^M \mathbf{v}_m B_{mi}=\sum_{m=1}^M \mathbf{v}_m \lambda_i \gamma_{mi}.$$
From this system we may find values of $\gamma_{mi}$ and $\lambda$. Note that $b_{mm}=e_m$ and $b_{ml}=b_{lm}^*$.  

\end{itemize}

\textit{Remark 2:}  Assume that 
$$\mathbf{R}=\mathbf{v}_1\mathbf{v}_1^H+\mathbf{v}_2\mathbf{v}_2^H+\mathbf{v}_3\mathbf{v}_3^H$$
and that $\mathbf{v}_3$ is not an independent vector, but a linear combination of $\mathbf{v}_1$ and $\mathbf{v}_2$, then 
 \begin{align*}
 \mathbf{q}_{1}&=\gamma_{11}\mathbf{v}_1+\gamma_{21}\mathbf{v}_2+\gamma_{31}\mathbf{v}_3\\
 \mathbf{q}_{2}&=\gamma_{12}\mathbf{v}_1+\gamma_{22}\mathbf{v}_2+\gamma_{32}\mathbf{v}_3.
 \end{align*}
reduces to
\begin{align*}
\mathbf{q}_{1}&=\beta_{11}\mathbf{v}_1+\beta_{21}\mathbf{v}_2\\
\mathbf{q}_{2}&=\beta_{12}\mathbf{v}_1+\beta_{22}\mathbf{v}_2.
\end{align*}
It means that a new dependent vector will not increase the dimensionality of the eigenvector space, and it will reduce to a linear combination of the independent vectors, with new coefficients. 

\textit{Remark 3:} If the vectors $\mathbf{v}_1$,  $\mathbf{v}_2$, \dots,  $\mathbf{v}_M$ are linear combinations of another set of independent vectors $\mathbf{w}_1$,  $\mathbf{w}_2$, \dots,  $\mathbf{w}_K$ then the eigenvectors as the linear combinations $\mathbf{v}_1$,  $\mathbf{v}_2$, \dots,  $\mathbf{v}_M$ are also the linear combinations of $\mathbf{w}_1$,  $\mathbf{w}_2$, \dots,  $\mathbf{w}_K$. For $M=2$, in the matrix form, for two vectors
\begin{gather*}
\begin{bmatrix}
\mathbf{q}_1 \\ \mathbf{q}_2
\end{bmatrix}
=\begin{bmatrix}
\gamma_{11} & \gamma_{21}\\
\gamma_{12} & \gamma_{22}
\end{bmatrix} \begin{bmatrix}
\mathbf{v}_1 \\ \mathbf{v}_2
\end{bmatrix}=\begin{bmatrix}
\gamma_{11} & \gamma_{21}\\
\gamma_{12} & \gamma_{22}
\end{bmatrix} \begin{bmatrix}
\xi_{11} & \xi_{21}\\
\xi_{12} & \xi_{22}
\end{bmatrix} \begin{bmatrix}
\mathbf{w}_1 \\ \mathbf{w}_2
\end{bmatrix} \\
=\begin{bmatrix}
\beta_{11} & \beta_{21}\\
\beta_{12} & \beta_{22}
\end{bmatrix}\begin{bmatrix}
\mathbf{w}_1 \\ \mathbf{w}_2
\end{bmatrix}.
\end{gather*}
Therefore, the eigenvectors $\mathbf{q}_m$ are linear combinations of $\mathbf{w}_1$,  $\mathbf{w}_2$, \dots,  $\mathbf{w}_K$.

\textit{Remark 4: } If the number of independent vectors $\mathbf{w}_1$,  $\mathbf{w}_2$, \dots,  $\mathbf{w}_K$ is $K$ and  $\mathbf{v}_1$,  $\mathbf{v}_2$, \dots,  $\mathbf{v}_S$, are their linear combinations with $S>K$, then only $K$ vectors $\mathbf{v}_i$ are linearly independent. This means that only $K$ eigenvectors can be formed in this basis.
      
\subsection{Eigenvectors as Linear Combinations of the Signal Components}
The previous remarks represent an analysis platform for our multivariate and multicomponent signal defined by (\ref{MVmcSIG}). The vectors that form the matrix $\mathbf{R}$ are formed as the following linear combinations of the signal component vectors
 $$ \mathbf{R}= \mathbf{X}_{sen}^H\mathbf{X}_{sen}=\mathbf{X}_{com}^H\mathbf{A}^H\mathbf{A}\mathbf{X}_{com},$$
 with the elements
\begin{gather*}
R(n_{1},n_{2})=\begin{bmatrix}
x_1^*(n_2), \,
x_2^*(n_2),\,
\dots, \,
x_P^*(n_2)
\end{bmatrix}\mathbf{A^H}\mathbf{A}
\begin{bmatrix}
x_1(n_1)\\
x_2(n_1)\\
\vdots \\
x_P(n_1)
\end{bmatrix}.
\end{gather*}
The eigenvalue decomposition is then given by
\begin{equation}
\mathbf{R=Q}\mathbf{\Lambda}\mathbf{Q}^{T}=\sum_{p=1}^{M}\lambda_{p}\mathbf{q}%
_{p}\mathbf{q}_{p}^{\ast}, \label{eig2}%
\end{equation}
where the eigenvectors are linear combinations of $\mathbf{x}^{(i)}$ and these components are linear combinations of the signal components. In other words
\begin{align}
\mathbf{q}_{1}&=\alpha_{11}\mathbf{x}_1+\alpha_{21}\mathbf{x}_2+ \dots +\alpha_{P1}\mathbf{x}_P \nonumber \\
\mathbf{q}_{2}&=\alpha_{12}\mathbf{x}_1+\alpha_{22}\mathbf{x}_2+ \dots +\alpha_{P2}\mathbf{x}_P \nonumber \\ & \qquad \qquad \qquad \vdots \nonumber \\
\mathbf{q}_{M}&=\alpha_{1M}\mathbf{x}_1+\alpha_{2M}\mathbf{x}_2+ \dots +\alpha_{PM}\mathbf{x}_P, \label{linqeq}
\end{align}
with $M=\min\{S,P\}$.

Consider the case when the signal components $\mathbf{x}_p(n)$ overlap in the frequency plane. In this case, the decomposition on the individual components is not possible using the state-of-art methods, except in cases of quite specific signal forms (such as linear frequency modulated signals, using chirplet transform, Radon transform or similar techniques \cite{dekompozicija_radon}, \cite{dekompozicija_cirplet}, or for sinusoidally modulated signals using inverse Radon transform, \cite{iradon1}, \cite{iradon2}). In general, these kinds of signals cannot be separated into individual components in the univariate case. However, the multivariate form of signals offers a possibility to decompose the components which overlap in the time-frequency plane.

\section {Decomposition Principle }
 \label{decomposition_principle}
We have concluded that the eigenvectors of matrix $\mathbf{R}$ are formed as $M=\min\{S,P\}$ linear combinations of the signal components in (\ref{linqeq}). Assume now that the number of sensors $S$ is such that $S\ge P$. Then there are $M=P$ independent linear relations for $P$ components.  We may conclude that, in principle, the signal component $\mathbf{x}_{p}$ can be also be written as linear combination of eigenvectors
$\mathbf{q}_p$
\begin{gather*}
\mathbf{x}_{p}=\eta_{1p}\mathbf{q}_{1}+\eta_{2p}\mathbf{q}_{2}+\dots+\eta_{Pp}\mathbf{q}_{P},
\end{gather*} 
with unknown weights $\eta_{ip}$.

We will consider signal with nonstationary components $\mathbf{x}_{p}$, $p=1,2,\dots,P$. Each component has a support in the time-frequency domain denoted by $\mathbb D_p$. For components with partial overlapping, both in time and frequency, the supports also partially overlap. The case with the complete overlapping of two supports is excluded from this analysis. Assume the notation such that $D_1 \le D_2 \le \dots \le D_P$, where $D_p$ is the area of the support $\mathbb D_p$. 

The aim of this paper is to decompose the original signal, using the eigenvectors, $\mathbf{q}_{p}$, $p=1,2,\dots,P$ of autocorrelation matrix $\mathbf{R}$, and to obtain  the individual signal components $\mathbf{x}_{p}$, $p=1,2,\dots,P$, by linearly combining the eigenvectors $\mathbf{q}_p$. To meet this aim, we will use time-frequency representations and corresponding concentration measures. Since the form of time-frequency representation is not crucial here, we will use the short-time Fourier transform (STFT),
\begin{equation}
STFT(n,k)=\sum_{m=0}^{S_w-1}w(m)x(n+m)e^{-j2 \pi mk /S_w}, \label{stftdef}
\end{equation} 
to measure the concentration of signals in the time-frequency domain, and the pseudo Wigner distribution  
\begin{equation}
WD(n,k)=\sum_{m=0}^{S_w-1}w(m)w(-m)x(n+m)x^*(n-m)e^{-j4 \pi \frac{mk} {S_w}},\label{wddef}
\end{equation} 
to visualize the results, that is, for a high resolution presentation of the initial signal, eigenvectors and the resulting signal components. Note that $w(n)$ denotes a window of length $S_{w}$ in (\ref{stftdef}) and (\ref{wddef}).
  
An $L_p$-norm based measure of the time-frequency concentration, with $0\le p \le 1$, will be used. It is originally introduced in \cite{XXX} as 
 \begin{gather}
 \mathcal{M}\left\{ STFT(n,k) \right\}
 =\Vert STFT(n,k)
 \Vert_p^p \\ =\sum_{n}\sum_{k} |STFT(n,k)|^{p}  =\sum_{n}\sum_{k} SPEC^{p/2}(n,k),
 \end{gather} 
 where  $SPEC(n,k)=|STFT(n,k)|^2$ is the spectrogram. 
  
In theory, a direct way to solve the problem of eigenvectors decomposition to the signal components would be to form a linear combination of the eigenvectors  
\begin{equation} 
\label{bete1}
\mathbf{y}=\beta_{1}\mathbf{q}_{1}+\beta_{2}\mathbf{q}_{2}+\dots+\beta_{P}\mathbf{q}_{P,}
\end{equation}
with varying coefficients $\beta_{p},~p=1,2,\dots,P$, keeping $\Vert\mathbf{y}\Vert_2=const.$, and to  use the zero-norm as the concentration measure. This norm would produce the area of the support for the analyzed signal. If all signal components are present in the signal $y(n)$, then its zero-norm would produce the area of $\mathbb D_1 \cup \mathbb D_2 \cup \dots \cup \mathbb D_P$. By changing the  coefficients $\beta_{p}$, the minimum value of the concentration measure is achieved when the coefficients $\beta_{p}$ are matched to the best concentrated signal component coefficients $\eta_{p1},~p=1,2,\dots,P$ with the smallest support area $D_1$  
$$[\eta_{11},\eta_{21},\dots,\eta_{P1}]=\arg \min_{\beta_{1},\dots,\beta_{P}}\Vert SPEC(n,k)
\Vert_0.$$ 
 If any two the smallest areas are equal, we will still find one of them. 
In practice, the norm-one of the STFT $\Vert STFT(n,k)
\Vert_1=\Vert SPEC(n,k)
\Vert_{1/2}$ could be used to achieve the robustness to noise
\begin{align}
[\eta_{11},\eta_{21},\dots,\eta_{P1}]=\arg \min_{\beta_{1},\dots,\beta_{P}}\Vert STFT(n,k)
\Vert_1.\label{konc}
\end{align}
 
Note that this minimization problem has several local minima as the coefficients $\beta_{p}$ in $\mathbf{y}=\beta_{1}\mathbf{q}_{1}+\beta_{2}\mathbf{q}_{2}+\dots+\beta_{P}\mathbf{q}_{P}$
which correspond to any signal component $\mathbf{x}_{p}$ will also produce a local minimum of the concentration measure, equal to the area of corresponding component support. In addition,  any linear combination of $K<P$ signal components  $\mathbf{x}_p$ will also produce a local minimum equal to the area of the union of the supports of included signal components. Note that if $P$ the lowest local minima correspond to $D_1$, $D_2$, \dots, $D_P$, then we can detect the coefficients for all signal components.

As several local minima exist, multicomponent decomposition should be performed iteratively. 
Initially, the matrix $\mathbf{R}$ with elements (\ref{relem}) is calculated as in (\ref{rdef}). Its eigen-decomposition produces eigenvectors $\mathbf{q}_p,~p=1,2,\dots,P$, and based on them, signal 
$$\mathbf{y}=\beta_{1}\mathbf{q}_{1}+\beta_{2}\mathbf{q}_{2}+\dots+\beta_{P}\mathbf{q}_P$$ is formed, with weighting coefficients $\beta_p,~p=1,2,\dots,P$, which are varied to solve the minimization problem
(\ref{konc}). The
STFT in (\ref{konc}) is calculated for the normalized signal $\mathbf{y}/\|\mathbf{y}\|_2^2={\mathbf{y}}/{{\sqrt{\|\sum_{p=1}^{P}\beta_p\mathbf{q}_p\|_2}}}$. Here, we can assume that the minimization (\ref{konc}) is performed by the direct search over the parameter space. 

Upon finding the concentration measure minimum, the eigenvector $\mathbf{q}_1$ is replaced with the signal $\mathbf{x}_{1}=\eta_{11}\mathbf{q}_{1}+\eta_{21}\mathbf{q}_{2}+\dots+\eta_{P1}\mathbf{q}_{P}$, formed using the weighting coefficients  corresponding to the minimum of concentration measure (\ref{konc}). Then, this signal is removed from the remaining eigenvectors, by removing its projection to these eigenvectors. In other words, the eigenvectors
 $\mathbf{q}_p,~p=2,3,\dots,P$,  are modified as follows: 
\begin{equation}
\mathbf{q}_p = \frac{\mathbf{q}_p - \mathbf{q}_1^H \mathbf{q}_p \mathbf{q}_1}{\sqrt{1-|\mathbf{q}_1^H \mathbf{q}_p|^2}},
\end{equation}
in order to ensure that $\mathbf{x}_1$ it is not detected again. 

This procedure is iterated $P$ times. This means that in the $i$-th iteration, based on eigenvectors $\mathbf{q}_p$ modified in the previous iteration, new signal
\begin{equation}
\mathbf{y}=\sum_{p=1}^P\beta_p\mathbf{q}_p,
\end{equation}
is formed. The weighting coefficients $\beta_p,~p=1,2,\dots,P$ are varied, to find the new set  $\eta_{1i},\eta_{2i},\dots,\eta_{Pi}$ which minimizes the concentration measure
$$[\eta_{1i},\eta_{2i},\dots,\eta_{Pi}]=\arg \min_{\beta_{1},\dots,\beta_{P}}\
	\Vert {STFT_y(n,k)}
	\Vert_1,$$
	of the spectrogram calculated for normalized current signal $\mathbf{y}/\|\mathbf{y}\|_2^2$. The $i$-th eigenvector is replaced by $\mathbf{x}_{i}=\eta_{1i}\mathbf{q}_{1}+\eta_{2i}\mathbf{q}_{2}+\dots+\eta_{Pi}\mathbf{q}_{P}$,  while the signal deflation \cite{SDP} is performed by subtracting the projection of the detected component from remaining eigenvectors $\mathbf{q}_p,~p=i+1,i+2,\dots,P$:
\begin{equation}
\mathbf{q}_p = \frac{\mathbf{q}_p - \mathbf{q}_i^H \mathbf{q}_p \mathbf{q}_i}{\sqrt{1-|\mathbf{q}_i^H \mathbf{q}_p|^2}}.
\end{equation}

The described procedure is repeated until there is no more updates of vectors $\mathbf{q}_p$. Vectors $\mathbf{q}_p$ are sorted according to their concentration measure, after each iteration. The iterative procedure is stopped when there is no updates of vectors $\mathbf{q}_p$.

The search in the space of parameters $\beta_1,\beta_2,\dots,\beta_P$, in order to minimize the measure $\mathcal{M}\left\{ STFT(n,k) \right\}=\Vert {STFT_y(n,k)}
\Vert_1$ can be performed directly, which is numerically inefficient, or by using more sophisticated methods, such as the iterative gradient minimization procedure presented in \cite{dekompozicija0}. Other global optimization methods, including heuristic algorithms - ant colony optimization \cite{ANT}, genetic algorithm, hill climbing \cite{HC}, simulated annealing \cite{SA}, and also, using some deterministic \cite{Deter} or stohastic procedures \cite{SPST,SPST1}, can be also used for the concentration measure minimization. However, this is out of the scope of this paper.

\begin{figure}[htbp]
	\includegraphics[scale=.85]{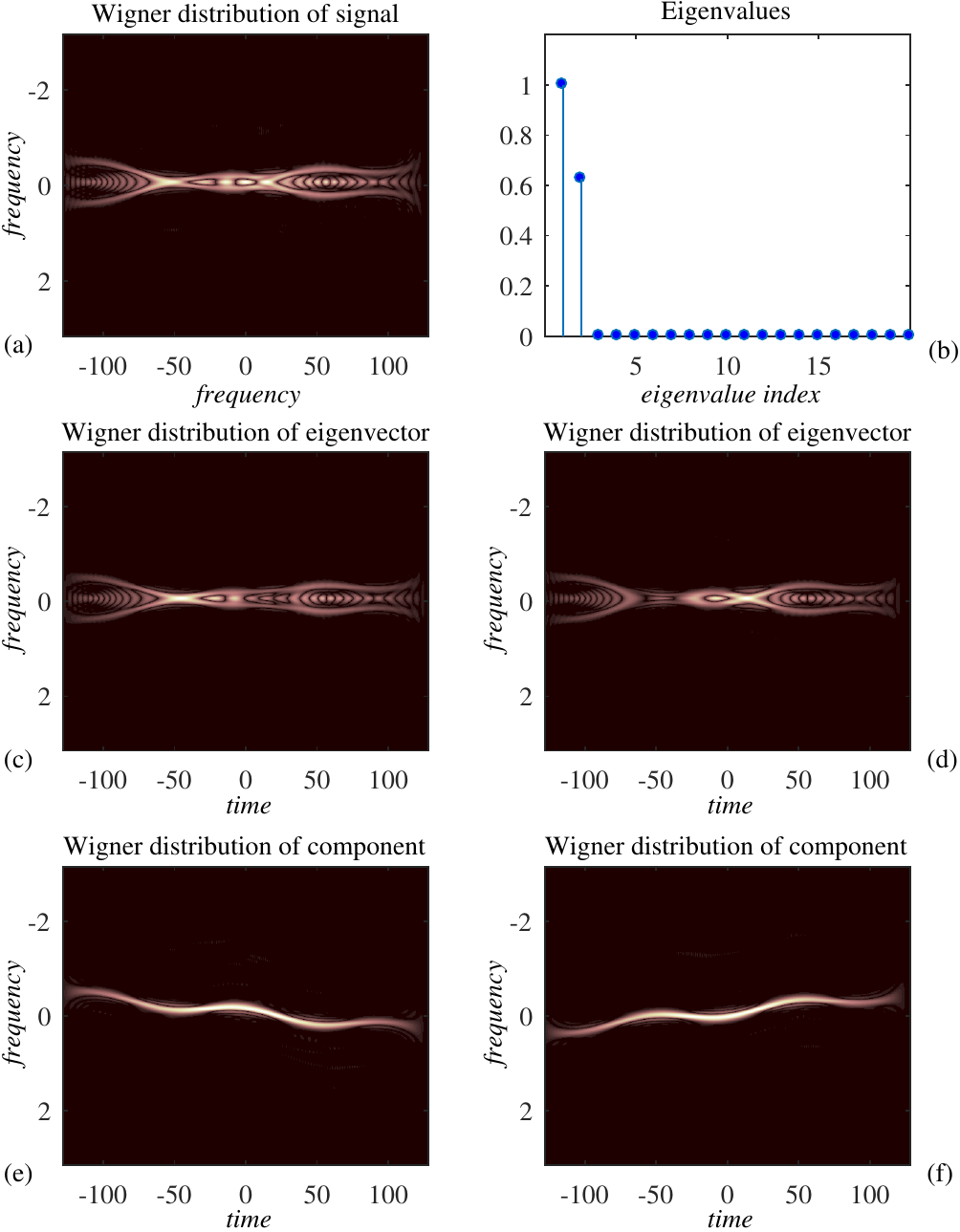}
	\caption{Signal Decomposition with a signal measured by $S=2$ sensors. Additive noise of the standard deviation $\sigma_{\epsilon}=0.01$ is present in the signal: (a) Time-frequency representation of the input signal. (b) Eigenvalues of the autocorrelation matrix $\mathbf{R}$. (c) Time-frequency representation of the first eigenvector. (d) Time-frequency representation of the second eigenvector. (e) Time-frequency representation of the reconstructed first signal component. (f) Time-frequency representation of the reconstructed second signal component. }
	\label{Algorith_1_Bohme_Fig1}
\end{figure}

\begin{figure}[htbp]
\includegraphics[scale=.85]{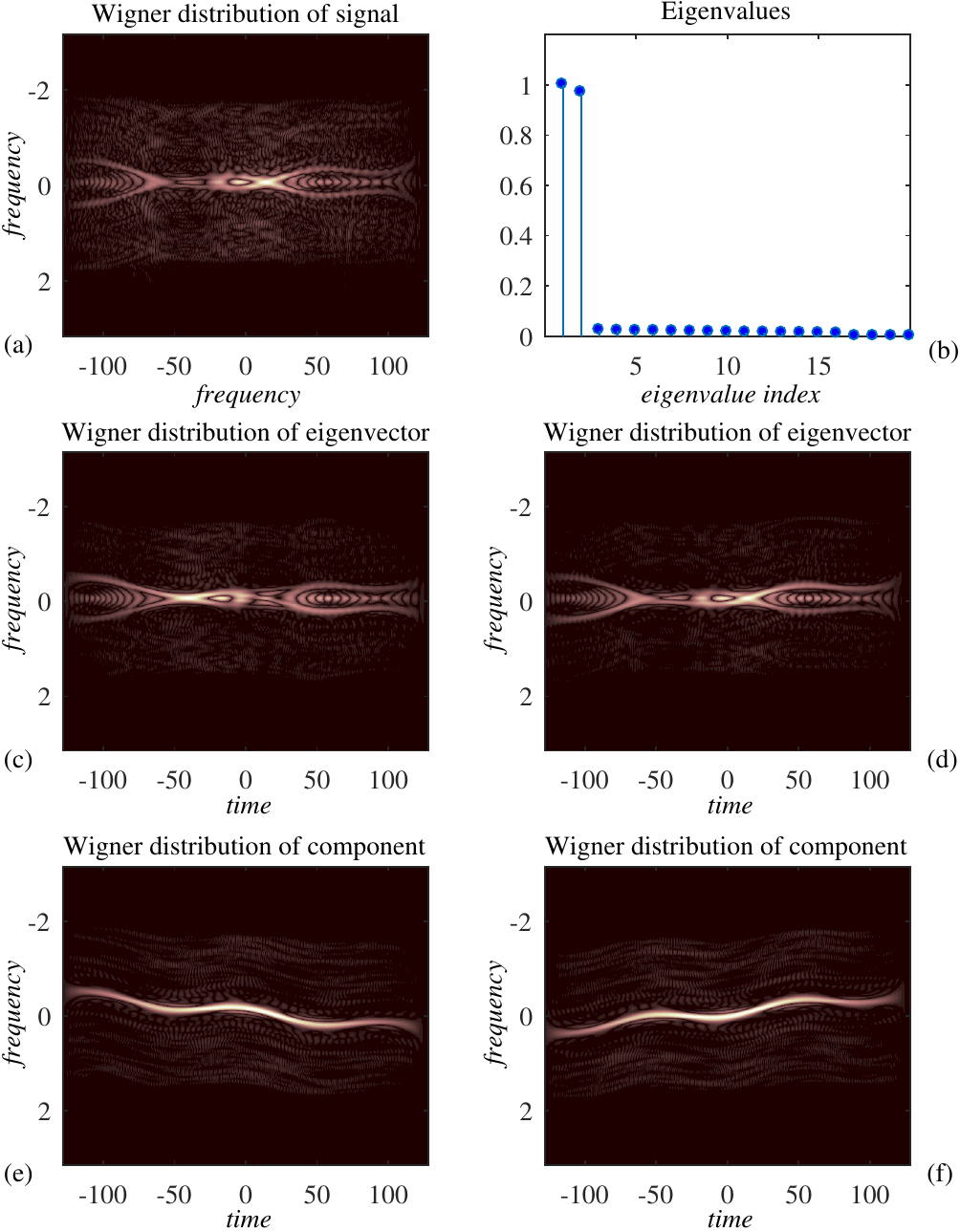}
\caption{Signal Decomposition with a signal from $S=16$ sensors. Additive noise of the standard deviation $\sigma_{\epsilon}=0.1$ is present in the signal: (a) Time-frequency representation of the input signal. (b) Eigenvalues of the autocorrelation matrix. (c) Time-frequency representation of the first eigenvector. (d) Time-frequency representation of the second eigenvector. (e) Time-frequency representation of the reconstructed first signal component. (f) Time-frequency representation of the reconstructed second signal component.}
\label{Algorith_1_Bohme_Fig2}
\end{figure}

\begin{figure}[htbp]
	\includegraphics[scale=.85]{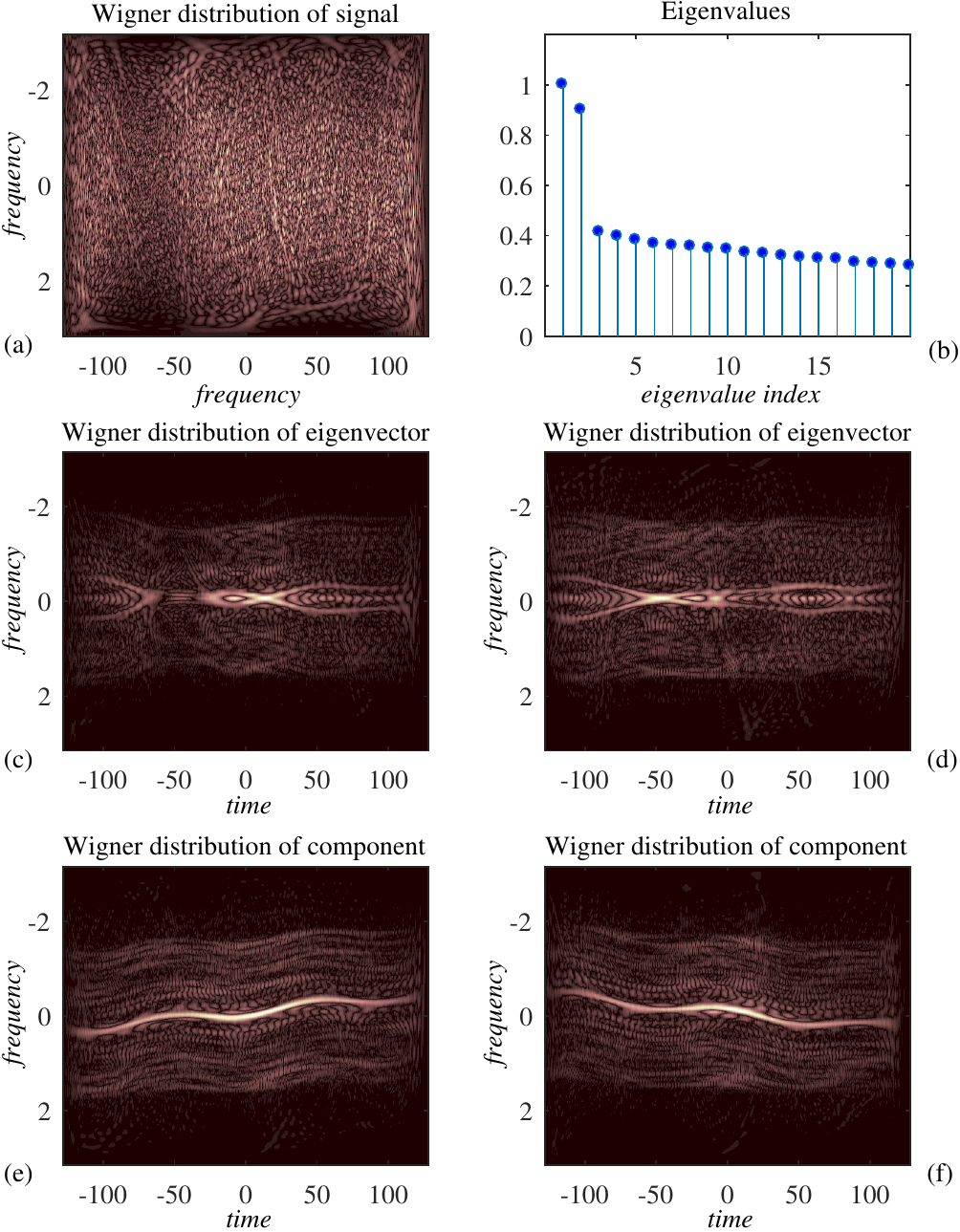}
	\caption{Signal Decomposition with a signal from $S=128$ sensors. Additive noise of the standard deviation $\sigma_{\epsilon}=1$ is present in the signal: (a) Time-frequency representation of the input signal. (b) Eigenvalues of the autocorrelation matrix. (c) Time-frequency representation of the first eigenvector. (d) Time-frequency representation of the second eigenvector. (e) Time-frequency representation of the reconstructed first signal component. (f) Time-frequency representation of the reconstructed second signal component.}
	\label{Algorith_1_Bohme_Fig3}
\end{figure}

\begin{figure*}[htbp]
	\centering
	\includegraphics[scale=.85]{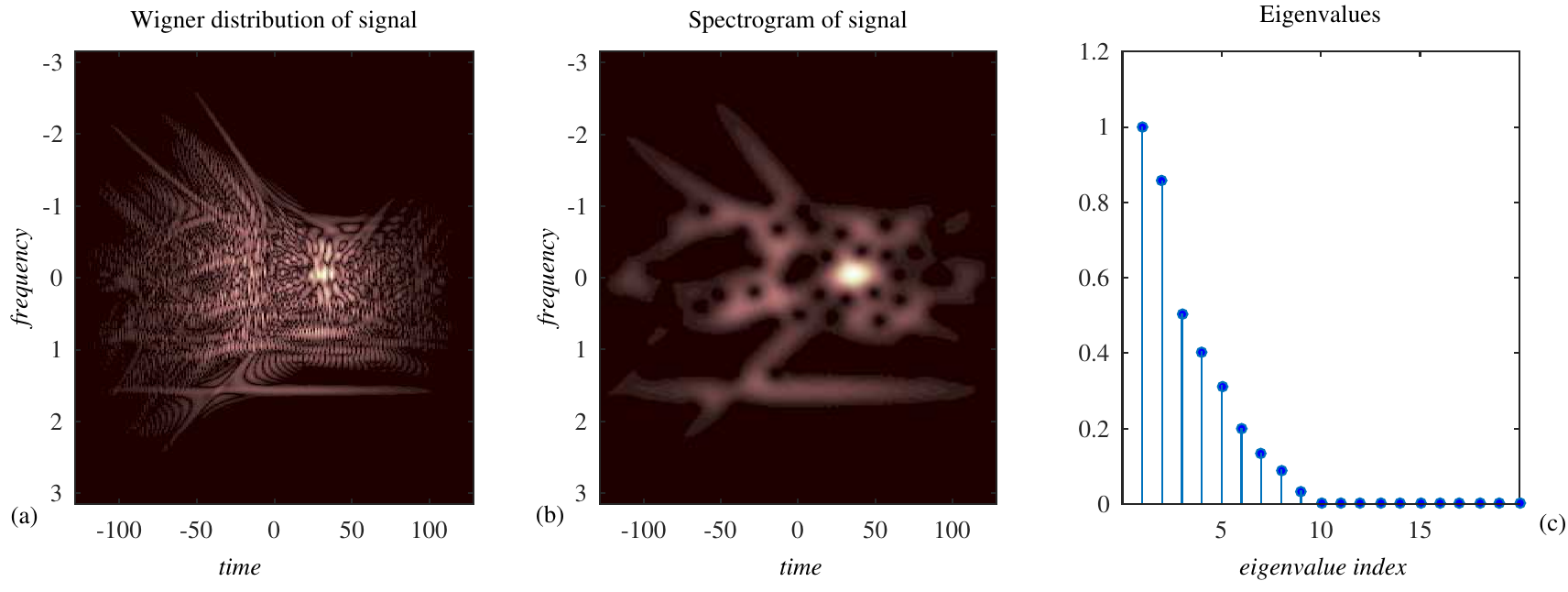}
	\caption{Time-frequency representation of a $P=9$ component signal using the Wigner distribution (left) and the spectrogram (middle), along with the eigenvectors of the autocorrelation matrix (right) obtained with $S=12$ sensors. Additive noise of standard deviation $\sigma_{\epsilon}=0.01$ is present in the input signal.}
	\label{Algorithm_1_Bohme_8comp_1}
\end{figure*}

\begin{figure*}[htbp]
		\centering
	\includegraphics[scale=.85]{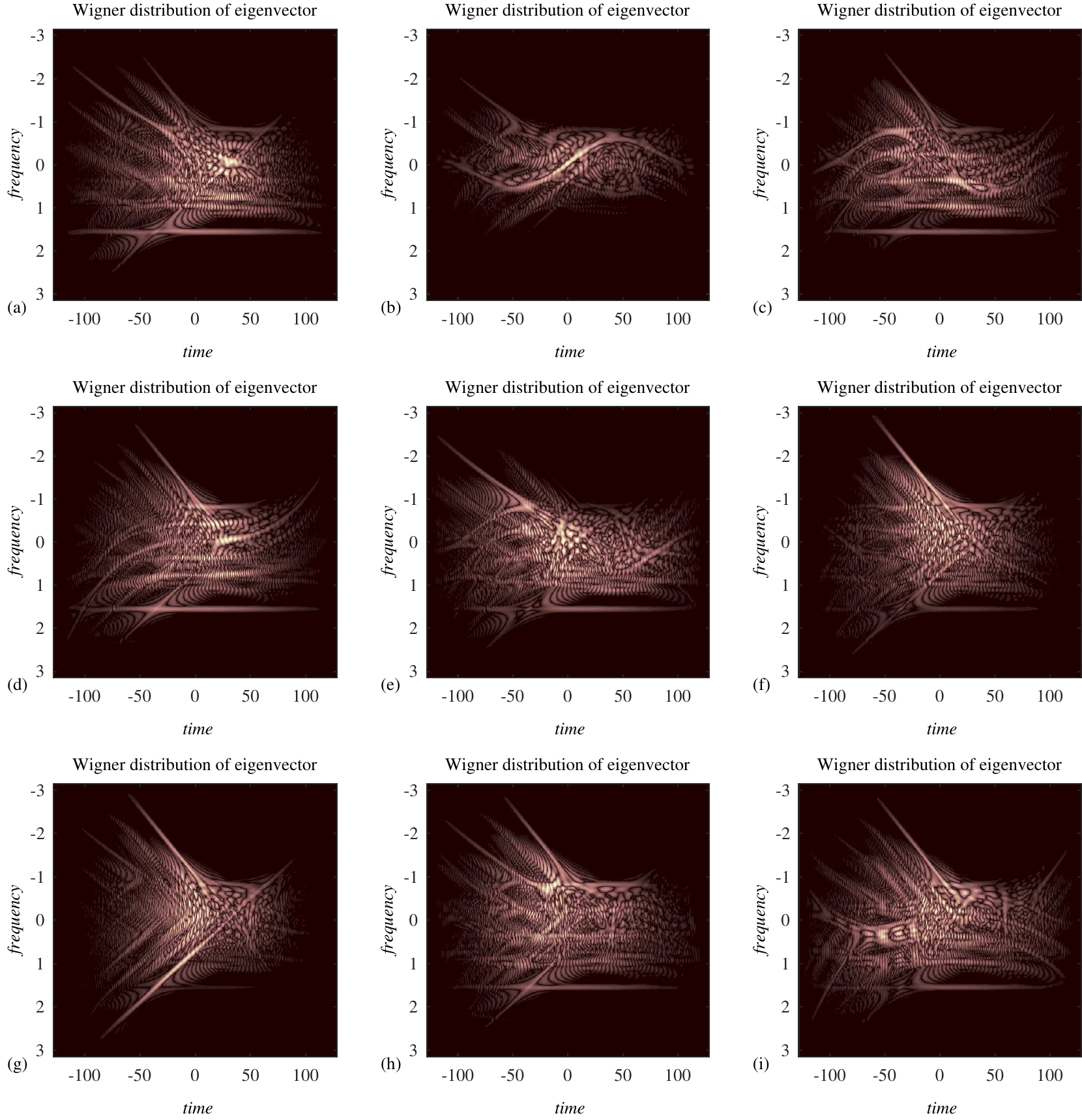}
	\caption{Time-frequency representation of $M=9$ eigenvectors of the autocorrelation matrix for the signal whose time-frequency representation is shown in Fig. \ref{Algorithm_1_Bohme_8comp_1}.}
		\label{Algorithm_1_Bohme_8comp_2}
\end{figure*}

\begin{figure*}[htbp]
		\centering
	\includegraphics[scale=.85]{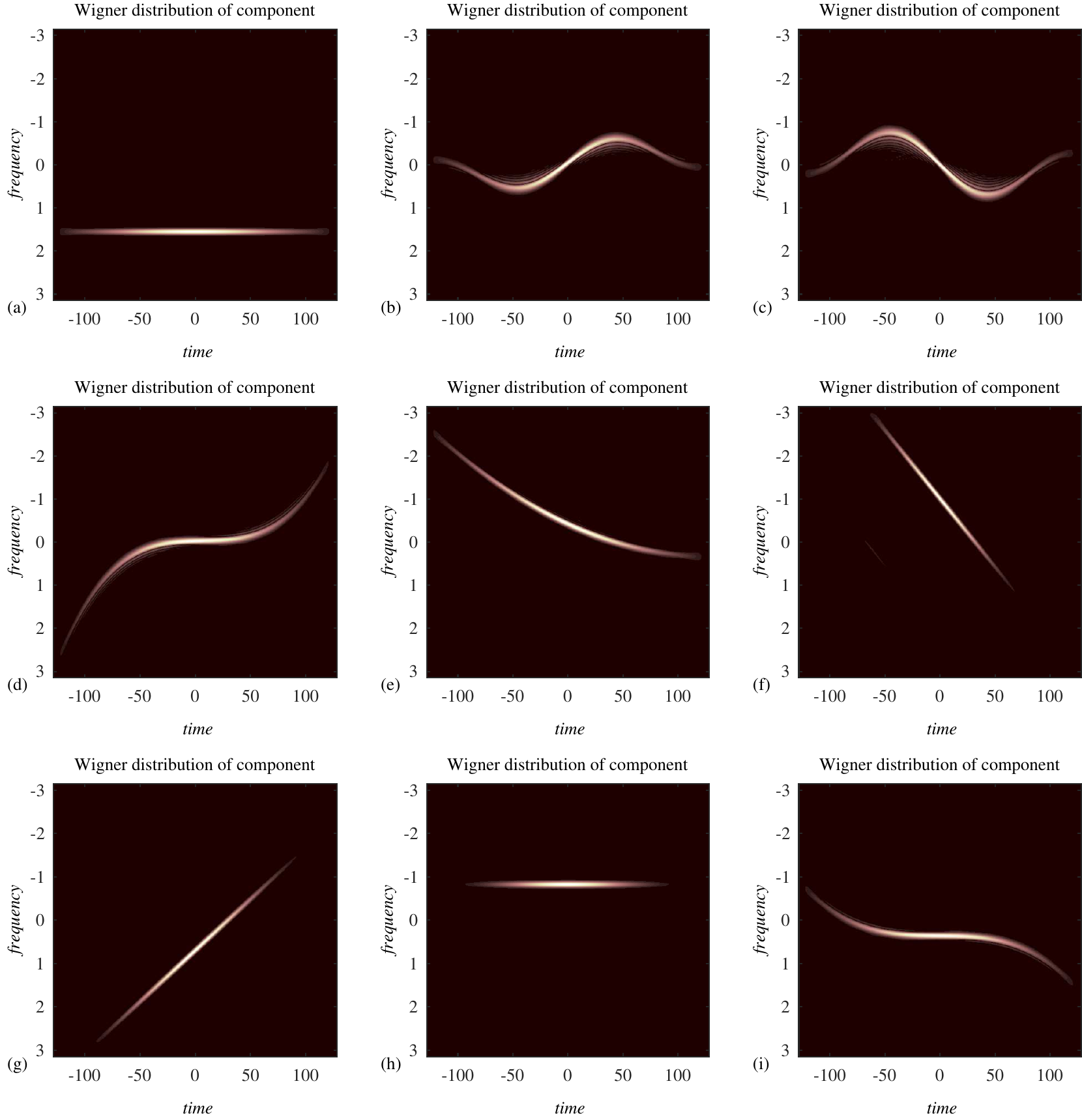}
	\caption{Time-frequency representation of $P=9$ signal components obtained using the presented algorithm and the eigenvectors from Fig.  \ref{Algorithm_1_Bohme_8comp_2} for the signal whose time-frequency representation is shown in Fig. \ref{Algorithm_1_Bohme_8comp_1}.}
		\label{Algorithm_1_Bohme_8comp_3}
\end{figure*}

\begin{figure*}[htbp]
	\centering
	\includegraphics[scale=.85]{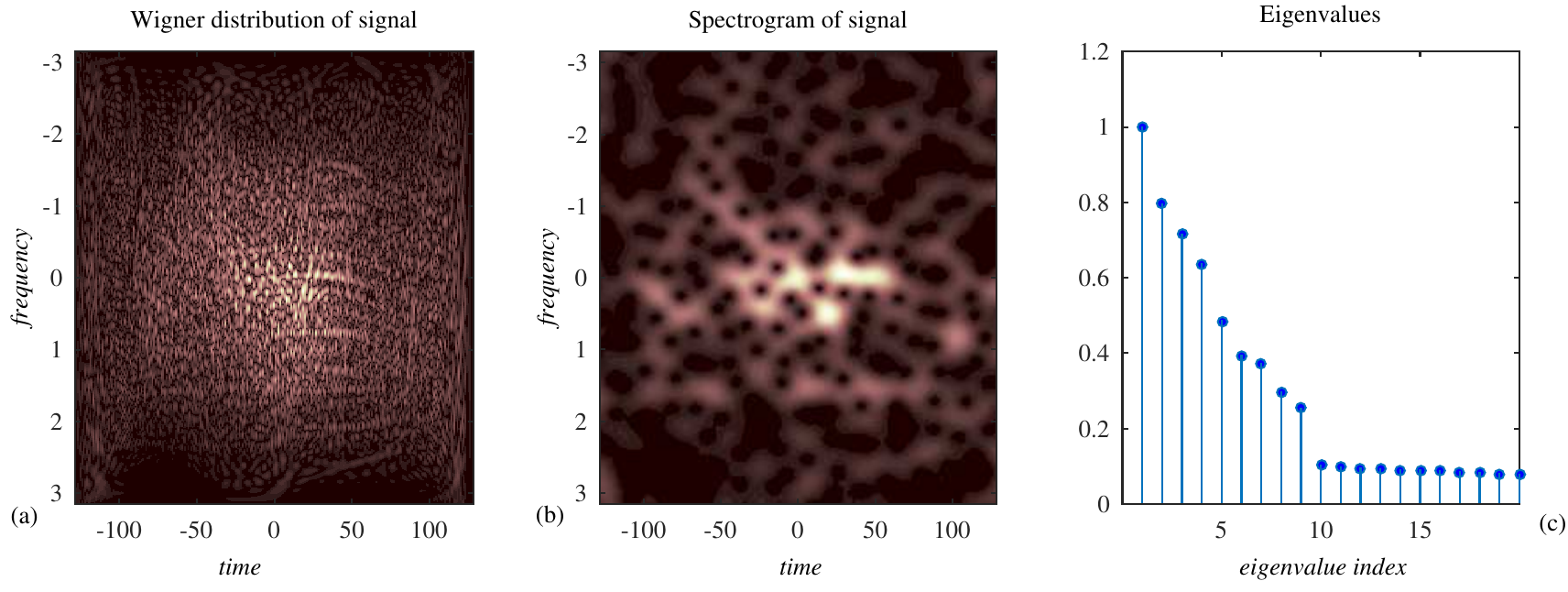}
	\caption{Time-frequency representation of a $P=9$ component signal using the Wigner distribution (left) and the spectrogram (middle), along with the eigenvectors of the autocorrelation matrix (right) obtained with $S=128$ sensors. Additive noise of standard deviation $\sigma_{\epsilon}=1$ is present in the input signal.}
	\label{Algorithm_1_Bohme_8comp_1Noise}
\end{figure*}

\begin{figure*}[htbp]
		\centering
	\includegraphics[scale=.85]{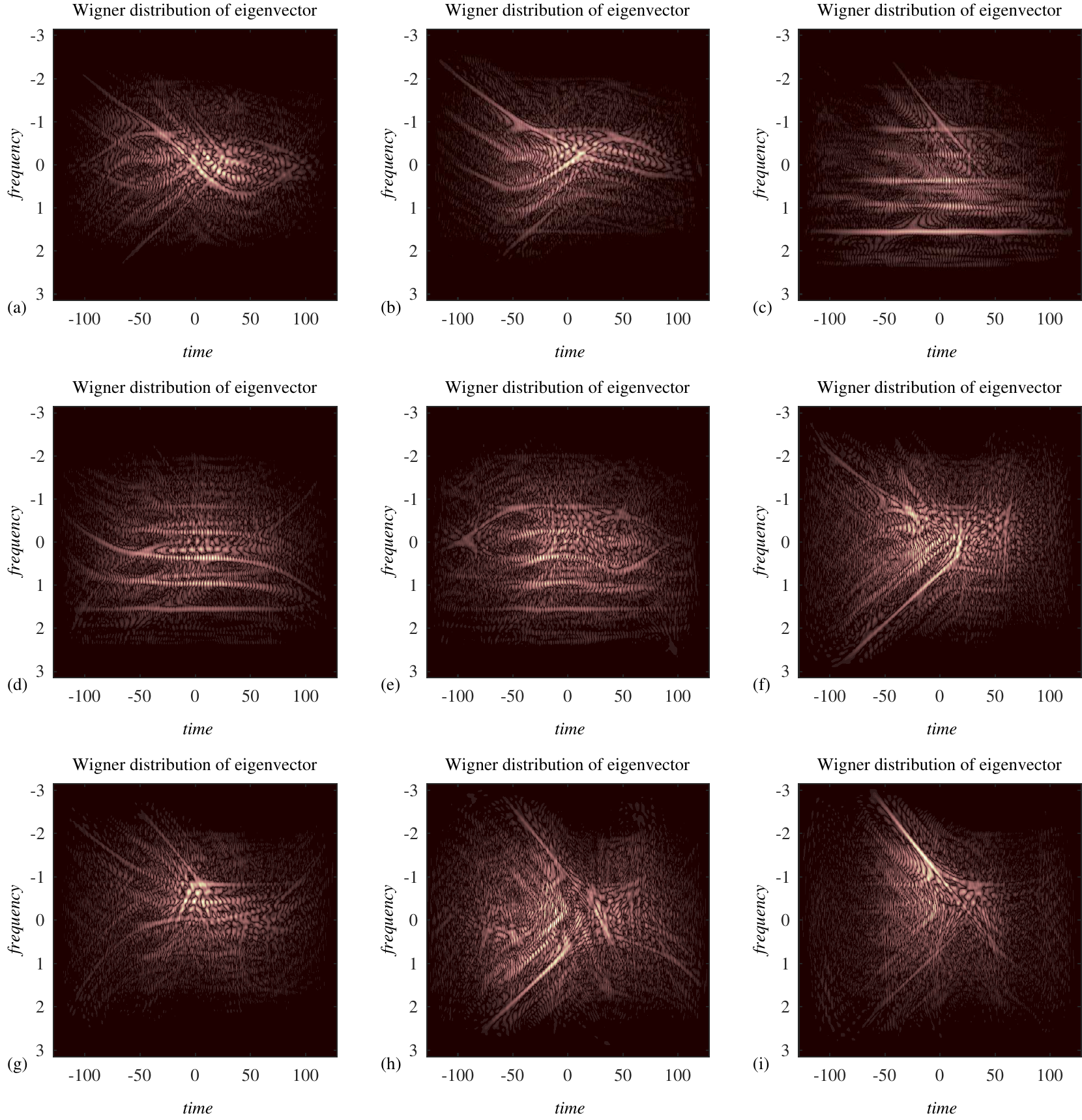}
	\caption{Time-frequency representation of $M=9$ eigenvectors of the autocorrelation matrix for the noisy signal whose time-frequency representation is shown in Fig. \ref{Algorithm_1_Bohme_8comp_1Noise}.}
		\label{Algorithm_1_Bohme_8comp_2Noise}
\end{figure*}

\begin{figure*}[htbp]
		\centering
	\includegraphics[scale=.85]{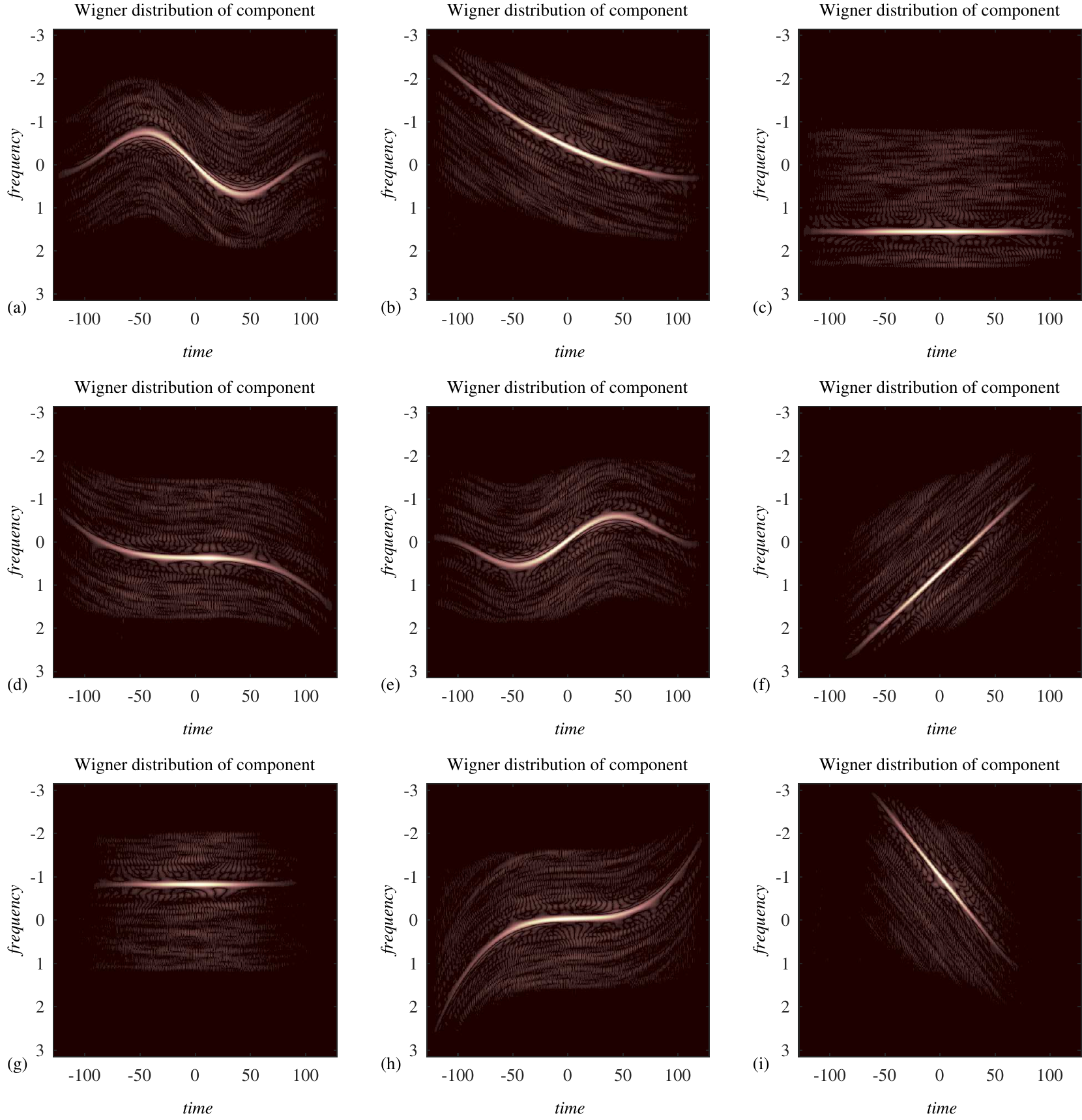}
	\caption{Time-frequency representation of $P=9$ signal components obtained using the presented algorithm and the eigenvectors from Fig.  \ref{Algorithm_1_Bohme_8comp_2Noise} for the noisy signal whose time-frequency representation is shown in Fig. \ref{Algorithm_1_Bohme_8comp_1Noise}.}
		\label{Algorithm_1_Bohme_8comp_3Noise}
\end{figure*}

\begin{figure*}[htbp]
		\centering
	\includegraphics[scale=.9]{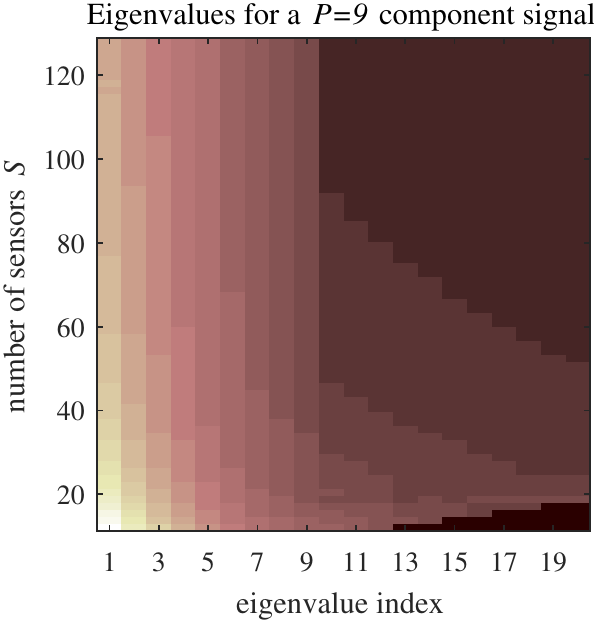} \hspace{5mm}
	\includegraphics[scale=.9]{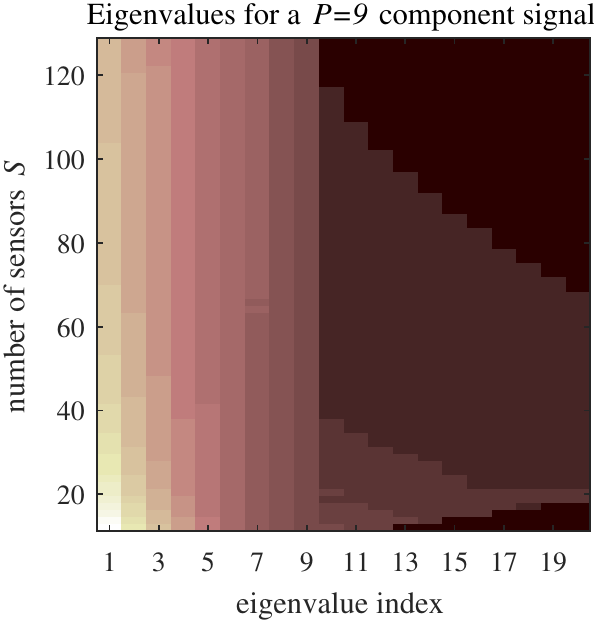}	
	\caption{Eigenvalues for a $P=9$ component noisy signal averaged over $1000$ random realizations, for two additive noise scenarios with $\sigma^2_{\varepsilon}=1$ (left)  and $\sigma^2_{\varepsilon}=1/2$ (right). The indicator of a successful reconstruction is the gap between the eigenvalues for the eigenvectors at the eigenvalue index equal to $P=9$ (representing the smallest energy of a combination of the signal components) and eigenvalue index equal to $P+1=10$ (representing the strongest background noise component). }
		\label{Variable_test0707}
\end{figure*}

\subsection{Specific Cases}
When the components do not overlap in the time-frequency plane, they are orthogonal. If the number of sensors is greater or equal to the number of components, $S\ge P$, then the components are equal  to the eigenvectors (up to their amplitudes) and the decomposition directly follows. In sense of the previous equations it means that we can use $b_{mn}=0$ for $m\ne n$. 

This problem can be solved even if single signal channel is available, $S=1$, by using time-frequency representation of the signal which produces the cross-terms free Wigner distribution -- the S-method, \cite{dekompozicija}. 

The case of combined $P_o$ overlapping and $P_n$ nonoverlapping components, $P=P_o+P_n$ can be solved with at least $S=P_o$ sensors, as shown in \cite{dekompozicija0}.

\section{Numerical Examples}
\label{examples}
This section supports the theory by on numerical examples. In Examples 1-3, a real-valued discrete-time bivariate signal with overlapping components is considered with various noise amounts (variances). This set of examples confirms the fact that in order to perform an efficient decomposition in noisy cases -- the number of channels should be increased, compared with the noiseless scenario. In Examples 4-5, a very complex signal of nine overlapping components is considered, corrupted with noise with two different levels. The analysis is concluded with a statistical test which will illustrate how the ability to separate the components depends on the noise variance and the number of channels.
 
\textbf{Example 1}: Consider a discrete-time bivariate signal of the form $\mathbf x(n)=[x_1(n),~x_2(n)]^T$. The minimum required number of sensors for this signal, $S=2$, is used. Signal from the channel $i$ is of the form
\begin{align}
x^{(i)}(n)=e^{-(n/128)^2}\cos \left(2\sin(5 \pi \frac{n}{N})-2 \pi \frac{n^2}{16N} +\varphi_{i}\right)\label{closesig}
\end{align}
 for $-128\leq n\leq128$ and $N=257$, as shown in Fig. \ref{Algorith_1_Bohme_Fig1}. The components of this signal are 
\begin{align}
x_{1,2}^{(i)}(n)=  e^{-(n/128)^2}e^{\pm j 2\sin  (5 \pi \frac{n}{N})-2k \pi \frac{n^2}{16N} +j\varphi_{i} }. \label{closesig2}
\end{align}
Time-frequency representation of this signal with two very close components is shown in Fig. \ref{Algorith_1_Bohme_Fig1}(a). The eigenvectors of the authocorrelation matrix indicate that there are two signal components, as shown in Fig. \ref{Algorith_1_Bohme_Fig1}(b). The two eigenvectors corresponding to the largest eigenvalues are presented in  Fig. \ref{Algorith_1_Bohme_Fig1}(c)-(d). These two eigenvectors are decomposed into two signal components with minimum concentration measures, as described in the previous section. The decomposition results are shown in Fig. \ref{Algorith_1_Bohme_Fig1}(e)-(f), and they fully correspond to the time-frequency representation of the individual signal components in (\ref{closesig2}).  

\textbf{Example 2}: The signal from Example 1 is corrupted by a moderate level of additive noise, to give $x^{(i)}(n)+\varepsilon^{(i)}(n)$. The standard deviation of noise is $\sigma_{\varepsilon}=0.1$. Here, we were not able to reconstruct the signal with the minimum number of sensors. To achieve a stable reconstruction, the number of sensors is increased to $S=16$. The time-frequency representation of the original noisy signal, eigenvalues, time-frequency representation of the eigenvectors, and the time-frequency representation of the obtained  signal components are shown in Fig. \ref{Algorith_1_Bohme_Fig2}.

\textbf{Example 3}: In this case the noise intensity is increased to the signal level using $\sigma_{\varepsilon}=1.$ 
To achieve robustness of the results, the number of sensors had to be increased. Noisy signal time-frequency representation, along with eigenvalues, time-frequency representation of the eigenvectors, and the time-frequency representation of the signal components are presented in Fig. \ref{Algorith_1_Bohme_Fig3}. 

\textbf{Example 4}: In this example, a signal with a large number of $P=9$ overlapped components is considered. The minimum number of sensors, required for the successful decomposition is $S=P=9$ in this case. Since a small noise is added, with  $\sigma_{\varepsilon}=0.01$, and the measured signal phases are random, the signal is reconstructed with a small margin in the number of sensors $S=12$.  From the time-frequency representation of components, presented in Fig. \ref{Algorithm_1_Bohme_8comp_1}(a)-(b), we can see that the components overlapping is significant.  Components cannot be recognized neither from the Wigner distribution nor from the spectrogram with an adjusted window. Their eigenvalues of the autocorrelation matrix are shown in Fig. \ref{Algorithm_1_Bohme_8comp_1}(c). The time-frequency representation of the strongest $9$ eigenvectors are presented in Fig. \ref{Algorithm_1_Bohme_8comp_2}. Using these eigenvectors the signal is decomposed into components, as shown in Fig. \ref{Algorithm_1_Bohme_8comp_3}.

\textbf{Example 5}: A noisy signal, as in Example 4, with $P=9$ components is analyzed here. In addition to the random different phases in each sensor, a random amplitude change is assumed as well. The coefficients in (\ref{MVmcSIG}) defined by  $a_{mp}=\alpha_{mp} e^{j\varphi_{mp}}$, are here used in the form $a_{mp}=(1+\nu_{mn})\alpha_{mp} e^{j\varphi_{mp}}$, where the random variable $\nu_{mn}$ assume the values within $-0.25\le \nu_{mn} \le 0.25$ and the variable $\varphi_{mp}$ is uniformly distributed over the interval from $0$ to $2\pi$.  The decomposition results are presented in the same way as in the previous figures. Time-frequency representations obtained using the Wigner distribution and the spectrogram are given in Fig. \ref{Algorithm_1_Bohme_8comp_1Noise}, along with the eigenvalues of the autocorrelation matrix. The time-frequency representations of the 9 strongest eigenvectors are shown in Fig. \ref{Algorithm_1_Bohme_8comp_2Noise}. The linear combinations of the eigenvectors are done according to the presented algorithm and the final results for the signal components can be seen in Fig. \ref{Algorithm_1_Bohme_8comp_3Noise}.

Finally, a statistical test is run for the noisy $P=9$ component signal from the last example. The eigenvalues are calculated in $1000$ random realizations and presented in Fig. \ref{Variable_test0707} for two values of the additive noise variance  $\sigma^2_{\varepsilon}=1$ and  $\sigma^2_{\varepsilon}=1/2.$ Ability to clearly separate the strongest $P=9$ eigenvectors, corresponding to the linear combinations of the signal components, from the background noise is a good indicator when the presented algorithm can successfully be applied. For the value of variance $\sigma^2_{\varepsilon}=1/2$ we can conclude that the number of sensors $S>50$ would be sufficient, while  the same separation gap is obtained for $S>100$ with $\sigma^2_{\varepsilon}=1.$ This indicator is verified against the reconstruction check for these scenarios. 

\section{Conclusion}
This work presents new contributions to the most challenging topic in multicomponent signal decomposition in the case of components for which the supports are overlapped in the time-frequency plane. The decomposition concepts have been investigated starting directly from the signal autocorrelation matrix of the input, whose eigenvectors can be linearly combined to form individual signal components. The decomposition procedure based on the presented theory has been evaluated through several numerical examples, and has conclusively verified the presented theory and the decomposition efficacy. For rigor, the robustness of the procedure, against the influence of an additive noise, has been studied from the perspective of the degrees of freedom, that is, number of sensors (channels) required to achieve a stable separation of signal components.

\end{document}